Published in *Light: Advanced Manufacturing* **5**, 8(2024). doi: 10.37188/lam.2024.009

# Low Bend Loss, High Index, Composite Morphology Ultra-fast Laser Written Waveguides for Photonic Integrated Circuits

**A. J. Ross-Adams[1], T. T. Fernandez[1], M. J. Withford[1] and S. Gross[1,2],**

[1] MQ Photonics, School of Mathematics and Physical Sciences, Macquarie University, 2109, Australia.

[2] School of Engineering, Macquarie University, 2109, Australia.

andrew.ross-adams@mq.edu.au

toney.teddyfernandez@unisa.edu.au

michael.withford@mq.edu.au

simon.gross@mq.edu.au

## Abstract

We demonstrate a novel, composite laser written 3D waveguide, fabricated in boro-aluminosilicate glass, with a refractive index contrast of $1.12 \times 10^{-2}$. The waveguide is fabricated using a multi-pass approach which leverages the respective refractive index modification mechanisms of both the thermal and athermal inscription regimes. We present the study and optimisation of inscription parameters for maximising positive refractive index change and ultimately demonstrate a dramatic advancement on the state of the art of bend losses in laser-written waveguides. The 1.0 dB $cm^{-1}$ bend loss cut-off radius is reduced from 10 mm to 4 mm, at a propagation wavelength of 1550 nm.

## Introduction

Ultrafast Laser Inscription (ULI) enables the parametric design and fabrication of Photonic Integrated Circuits (PICs). The technique was first discovered in the 1996 by Davis *et al.* and utilises a focused femtosecond laser to induce non-linear photoionisation in transparent dielectric media such as glasses, crystals and polymers, resulting in permanent, highly localised refractive index modifications[1]. In contrast to traditional planar lightwave circuits (PLCs), ULI eschews the photomask requirement inherent to conventional photolithographic methods, enabling rapid prototyping. As with PLCs, a wide range of functional circuit elements can be fabricated, such as power splitting Y-branches[2], directional couplers[3], and multimode interference couplers[4]. Furthermore, it is a non-planar process capable of true 3D device geometries, and a diverse range of modification regimes, including optical waveguides, microfluidic channels, micro-voids, trenches and cantilevers[5]. Access to three dimensions enables more complex structures such as photonic lanterns[6] and multicore fibre fan-outs capable of mapping to arbitrary core layouts[7]. The ULI platform supports applications including astrophotonics[8], on-chip quantum[9], on-chip cubesat swarm navigation[10], lab-on-a-chip[11], and space division multiplexing technologies for increasing telecommunications bandwidth and connection density[12]. In particular, there is vested interest in the development of Universal Photonics Processors (UPPs) which are capable of programmable signal routing[13].

The loss performance of a PIC is defined by its baseline linear propagation loss, bend losses and fibre coupling losses. Underpinning these loss components is the refractive index delta between the core and the substrate bulk material, which governs optical mode confinement. Many approaches to optimising index contrast have been reported over the years, such as tuning the substrate's material composition[14], waveguide-air interfaces[15], waveguide compression using damage-induced stress fields[16], exploiting electronic resonances using laser-induced band-gap shifts[17] and thermal annealing[18]. ULI waveguides are most commonly fabricated in fused silica, soda-lime and especially, boro-aluminosilicate display glasses originating from the consumer electronics industry, owing to their low cost, mechanical robustness and the favourable waveguide morphologies possible with the femtosecond laser writing technique. Waveguide index contrasts in these materials are generally measured in the $10^{-3}$ range, which is comparable to that of standard 1550 nm single-mode optical fibres (approximately $0.5 \times 10^{-3}$ for Corning SMF28 Ultra, guiding a 10.4 μm diameter mode-field). It should be noted that higher index contrasts in the $10^{-2}$ range are attainable, however this often sacrifices circularity and smoothness of the modification morphology[14]. The highest index contrast reported in silica glass is $2.2 \times 10^{-2}$ (ref. 19), though a higher delta of $4.5 \times 10^{-2}$ has since been reported in a bismuth germanate crystal[20]. Though germanium doped silica PLCs offer higher index contrast, typically on the order of 2% (ref. 21), the ULI platform offers fine control over waveguide parameters, enabled by the laser writing method. By appropriately tuning inscription parameters, one may control waveguide width, cross-sectional shape and index contrast, allowing for advanced waveguide functionality such as arbitrary polarisation control for on-chip beam rotation[22].

Scaling the complexity of laser written PICs, especially for future telecommunications applications such as UPPs and multicore fiber fan-outs, requires reducing the tolerable, minimum radius of waveguide curvature. Tighter bends allow for more compact waveguide routing within a chip, and hence, smaller chips with increased component density and lower insertion loss. The present state of the art in this regard was reported by *Lee et al.* in 2021, where a precisely controlled laser-induced crack at the edge of the waveguide was used to reduce the minimum bend radius to 10 mm at a wavelength of 1550 nm[23].

We propose a novel, non-damaging, multi-pass fabrication technique in boro-aluminosilicate glass (Corning Eagle XG), for increasing refractive index contrast, which involves two sequential fabrication steps. The first pass inscribes a smooth index change thermal waveguide with a laser repetition rate of 5.1 MHz -- this type of modification is henceforth referred to as a thermal waveguide. The second pass overwrites the core with a multi-scan athermal modification, with a repetition rate of 150 KHz, in order to enhance the index contrast, and thus, reduce bend losses. The thermal regime relies on photoionisation induced ion migration, leading to compositional change, whilst the athermal regime involves the formation of non-bridging oxygen hole centres (among other defects), to effect a positive refractive index change[24,25]. In this manuscript, we show that these two mechanisms can be triggered sequentially, for improved index contrast. We present the process of optimising multi-pass fabrication with respect to pulse energy, feedrate, pulses per mm and spherical aberration, as well as characterisation methods and the procedure for achieving high accuracy alignment to the waveguide core on the second the pass.

## Results

A chip was fabricated containing two blocks of thermal waveguides, as illustrated in Figure 2. (a) The athermal overwrite segments feature 3 mm long adiabatic tapered terminations, where the pulse energy is ramped in a linear fashion from 0 – 200 nJ. to ensure a smooth transition. The chip contains two blocks of waveguides, corresponding to two respective characterisation experiments. All experimental measurements, hereon, were performed using robotic 6-axis active alignment for coupling the launch and collector single-mode fibres to the chip.

### Block 1 – Linear Propagation Loss

This block is comprised of straight waveguides which feature multi-pass segments of incrementally shortened length, ranging from 8 mm to 2 mm, ending with an unmodified reference waveguide; this is analogous to a cutback measurement. The linear propagation loss of the composite waveguide was obtained by measuring the insertion loss of all waveguides in this block and extracting the slope of the linear relationship between insertion loss and composite segment length. The y-axis intercept of the loss curve represents the aggregate contribution of coupling losses and the scattering losses associated with the athermal termination tapers. A comparison of measured mode-field profiles is shown in Figure 1. Coupling losses were obtained from the overlap integral of the imaged fibre and thermal waveguide modes, hence, from the subtraction, the termination taper losses were calculated. The linear propagation losses of the thermal waveguides are measured simply by taking the average of the insertion loss of the 4 reference waveguides, subtracting the coupling loss and dividing by the length of the chip. These data are summarised in Table 1.

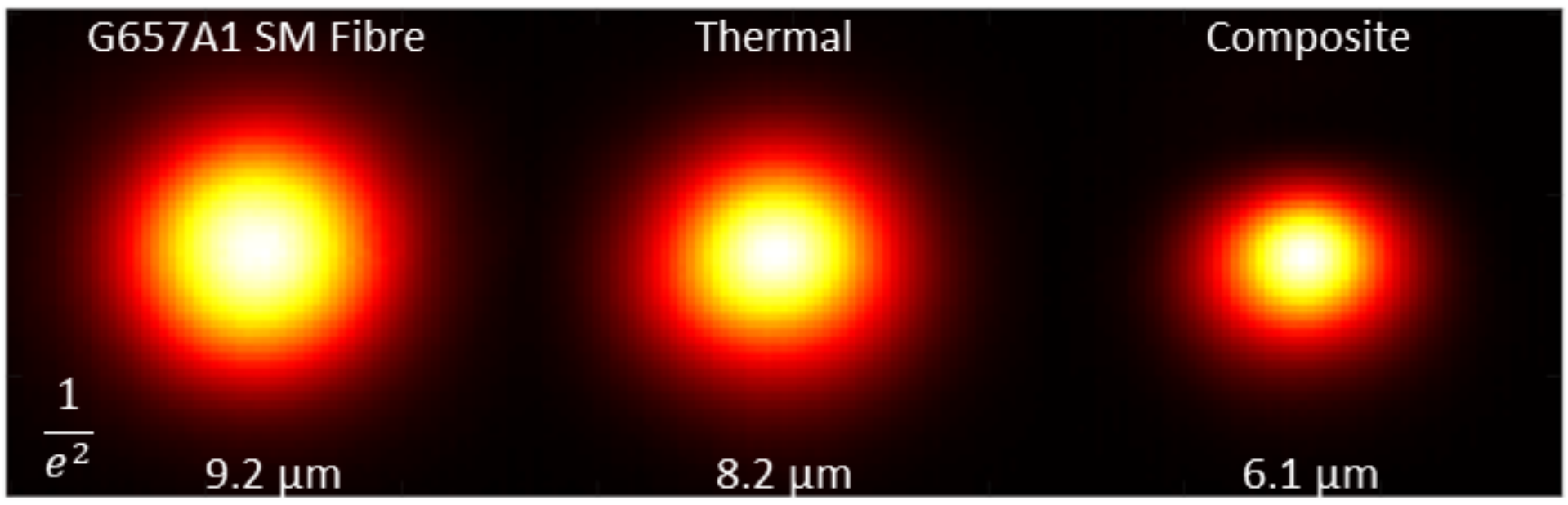

**Figure 1. Measured mode-fields of a standard single-mode fibre, a thermal waveguide and a composite waveguide guiding at a wavelength of 1550 nm, imaged with an InGaAs SWIR camera.**

| **Loss Component** | **Magnitude** | **Unit** |
|---|---|---|
| Linear Propagation - Thermal | 0.18 | dB $cm^{-1}$ |
| Linear Propagation - Composite | 0.50 | dB $cm^{-1}$ |
| Coupling Mode-Field Mismatch | 0.06 | dB |
| Athermal Termination | 0.18 | dB |

**Table 1. Waveguide linear loss components**.

The excess loss of the athermal termination tapers is due to localized glass damage in the low power region of the taper. This loss component also varies between 0.10 and 0.18 dB. This variance arises from high sensitivity to the precise transverse positioning of these damage effects, relative to the waveguide core. For rigor, the worst-case value of 0.18 dB is used for subsequent simulation.

## Block 2 – Bend Loss Study

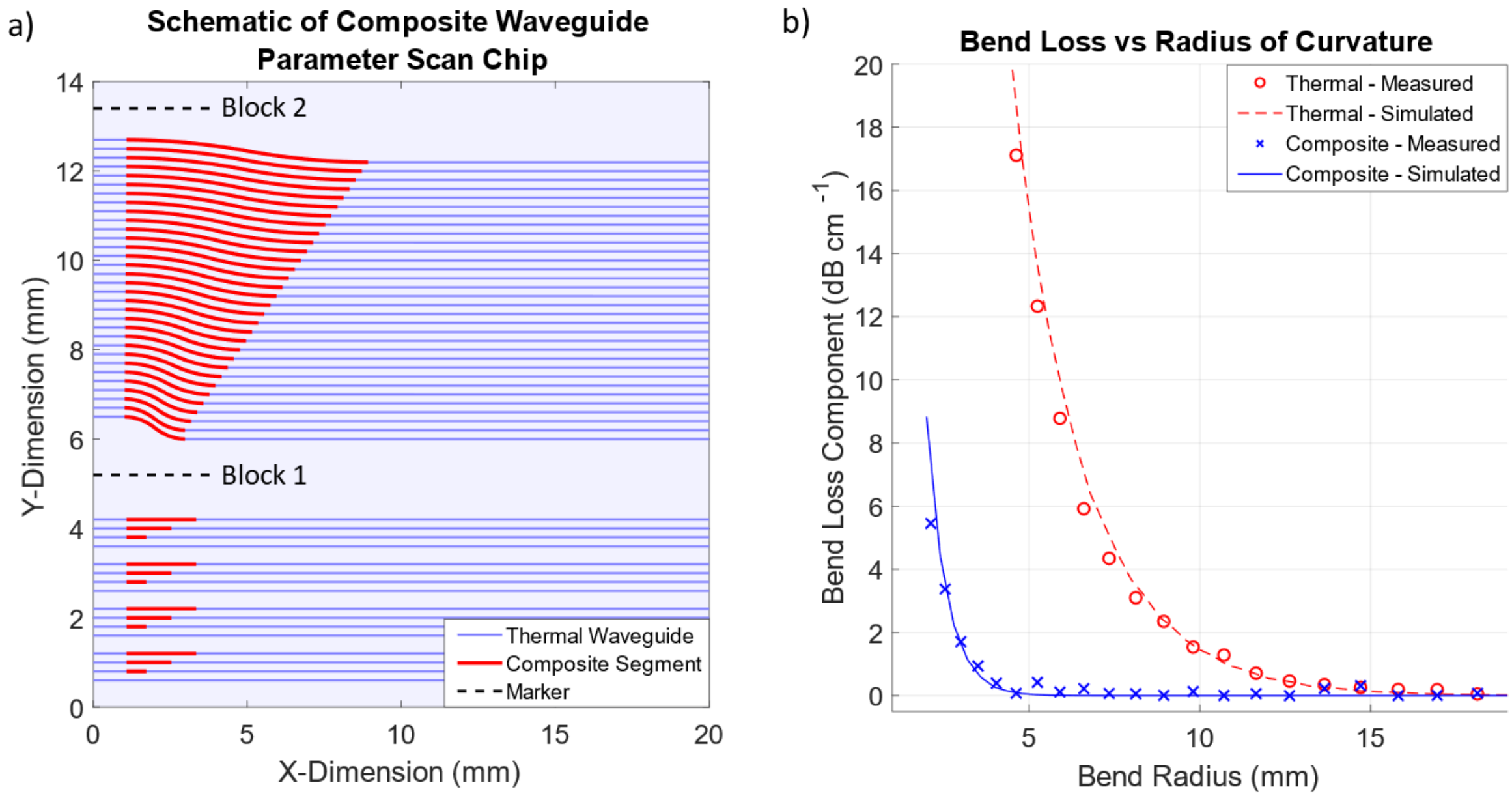

**Figure 2. a) Schematic diagram of the multi-pass experiment chip. The upper block of waveguides iterates the minimum radius of curvature across a series of circular arc s-bends, in order to characterise the evolution of the bend-loss component. The lower block is a set of straight waveguides featuring composite segments of decreasing length, in order to measure linear propagation losses. The red lines denote regions of composite modification, where the waveguide core has been overwritten with an athermal modification. b) Plot showing the simulation and measurement of the bend-loss component (i.e. excluding linear propagation and coupling losses) for both, conventional thermal waveguides and the novel composite waveguide for progressively smaller radii of curvature at 1550 nm wavelength.**

The second block contains sets of waveguides featuring s-bends based on two concatenated circular arcs with a total y-dimension side-step of 500 µm over a prescribed length L in the x-dimension (see Figure 2. a)), which is varied from 8 mm to 2 mm in 0.2 mm increments. Circular arcs were selected since they feature constant curvature, the radius of which, scales proportionately to L, equating to a bend radius scan spanning 32.1 mm to 2.1 mm. For each value of L, two waveguides are studied. The first is a purely thermal modification which serves as a reference. The second waveguide features a composite modification segment spanning the length of the s-bend and a 3 mm adiabatic transition region at each end. The total insertion loss was decomposed in order to extract the underlying pure bend-loss component, which is plotted as a function of bend radius in Figure 2. b). The composite waveguide offers clear advantages in tight bending scenarios, which are sufficient to outweigh the associated loss penalties caused by scattering and the taper losses. Subtracting the coupling losses and the contributions of the straight waveguide lead-in segments, and considering the insertion loss of only the s-bends, the abscissa at which the thermal and composite regimes have equal insertion loss occurs at a bend radius of 9.8 mm. For a 500 μm side-step, the measured abscissa corresponding to a 1.0 dB cut-off threshold for s-bend insertion loss occurs at a bend radius of 3.0 mm for the composite waveguide and 9.15 mm for the thermal waveguide, In other words, the minimum x-dimension length required to achieve 1.0 dB insertion loss s-bend with a 500 μm side-step was reduced from approximately 4.2 mm to 2.4 mm, a reduction of 43%. Next, considering specifically the extracted bend-loss component, the 1.0 dB $cm^{-1}$ cut-off occurs at a bend radius of 11.0 mm and 3.5 mm, for the thermal and composite waveguides, respectively, a 69% improvement which compares favourably with the state-of-the-art previously reported at 10 mm, using an integrated microcrack technique[22].

## Extracting Refractive Index Contrast

It was not feasible to directly measure the transverse refractive index profile of the composite waveguide, due to limitations associated with the respective measurement techniques. Measurement of thermal waveguides is achieved using the refracted near-field technique (RNF)[26]; however this method is unsuitable for studying the athermal modifications due to the presence of highly scattering microvoids in the region of negative index change. Conversely, quantitative phase imaging (QPI)[27], which measures optical pathlength differences, is unsuitable for studying thermal waveguides due to the requirement for a short modification length. This is problematic because in the thermal regime it is not possible to fabricate a perfect waveguide termination, instead, the transient dissipation of the plasma results in rounded off and tapered structures, for which no valid measurement can be recorded.

Importantly, both techniques measure the refractive index distribution in the visible range (632 nm for Rinck RNF profiler, 600 nm for the Phasics QPI camera), while the structures under test are intended to guide at 1550 nm. It was found that this leads to an overestimation of refractive index change. Hence, it was necessary to extract the refractive index contrast via comparative study of the experimentally measured bend losses and simulation (RSOFT FEMsim), based on the finite element method. This is achieved by stepping the refractive index delta of the simulation until agreement is reached by the simulated and experimentally measured bend loss curves.

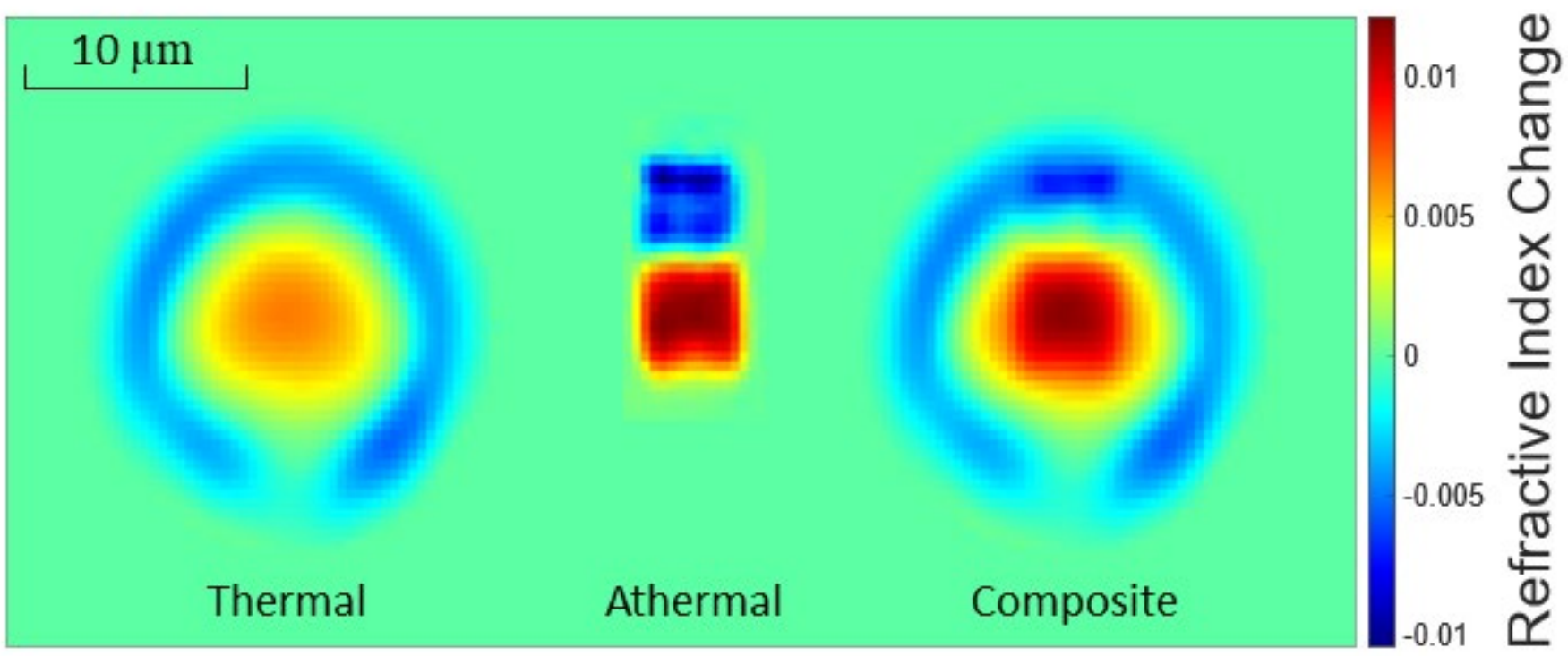

**Figure 3. Transverse refractive index profiles measured using the refractive near-field technique and quantitative phase imaging, respectively, for the thermal and athermal modifications. Also shown is the synthetically derived composite profile used for simulation.**

Bend loss is extremely sensitive to refractive index contrast. Consequently, the bend loss curve represents a very robust tool for extracting the index contrast. For example, the peak positive index change of the thermal waveguide profile is $8.5 \times 10^{-3}$ according to the RNF measurement. However, from the bend-loss curve, the true value is revealed to be $6.5 \times 10^{-3}$, a 30% overestimation. Since the composite index profile cannot be measured directly with either of the two techniques, a synthetic approximation was crafted by additively combining the respective profiles of the two underlying modifications in MATLAB. The refractive index addition is not one-to-one, but instead equivalent to the thermal waveguide plus approximately 44% of the athermal index change. This value was determined via iterative simulation where the contribution of the athermal modification is progressively increased until good agreement with measured data occurred. Further study is required to determine the physical explanation for the incomplete addition; we note that the thermal inscription pass has changed the structure and composition of the glass in the focal volume and the incident beam is being focused through the existing waveguide which influences the focal intensity distribution. The final value for the real positive index contrast (relative to the bulk material) was found to be $1.12 \times 10^{-2}$, or 0.8%. The respective transverse refractive index profiles are shown in Figure 3. The polarization dependence of the composite waveguides has not been explicitly studied yet. However, the birefringence of athermal multiscan waveguides and annealed thermal waveguides, have been previously measured to be on the order of $2.5 \times 10^{-4}$ and $1.2 \times 10^{-6}$, respectively, in boro-aluminosilicate glass[28,29].

**Discussion**

S-bends are essential for any photonic integrated circuit, especially UPPs. To fully contextualise and visualise the merit of the composite waveguide approach, it is necessary to consider other s-bend geometries, since circular arcs - by their nature - feature constant curvature. Hence, the bend-loss component is constant across the entire span of the s-bend. Consequently, it is often preferable to design s-bends based on raised-sine and cosine curves, which allow for gradual transitions, while reducing the path length over which, tight bending occurs. Thus, these three curve geometries form the foundational design "tool-box" for PIC routing. Bezier[30] and Euler[31] curves are also widely used for routing in more complex device geometries, though these are not being considered here. By fitting an exponential model to the measured bend-loss as a function of curvature, the insertion loss characteristics of an arbitrary s-bend can be extrapolated for any given s-bend length and side-step, and for any given curve equation. Side-step refers to the delta between the respective y-dimension (horizontal) positions of the s-bend start and termination points. Figure 4. shows comparative models of the insertion loss of arbitrary s-bends based on circular arcs and cosine bends, respectively, with, and without, the application of the composite waveguide technique. The insertion loss is calculated by applying the aforementioned bend loss model and the measured linear propagation loss to the total path length of the hypothetical s-bend, and then applying the addition of the athermal termination taper losses. The white regions represent the span of the parameter space, for which a 1.0 dB insertion loss cut-off criterion is not satisfied, and thus visually highlight the limitations of the respective fabrication regimes. Figure 4. a) shows the overlapping insertion loss surfaces for cosine and arc based s-bends written purely in the thermal regime. Comparing a) to b), the engineering advantage of the low composite waveguide bend losses is clearly illustrated by the dramatic reduction in the area of the region for which it is not possible to produce a sub 1.0 dB insertion loss s-bend. The most significant limiting factor constraining the efficacy of the composite waveguide tool-box is the athermal termination taper loss. This is especially important for cosine and raised sine s-bend, which feature two separate regions of high curvature separated by a region of zero curvature, hence, the taper loss penalty is incurred 4 times. In the model presented, the composite cosine s-bend is defined as a thermal waveguide, featuring the athermal overwrite modification only in regions where the radius of curvature is less than 10 mm. The taper losses may readily be reduced from 0.19 dB to 0.05 - 0.10 dB, by optimising the taper length and width parameters, and by progressively increasing the scan-line pitch to push the damage region outside of the core. Referring, again, to Figure 4. b), the taper losses need only be reduced below 0.15 dB for the composite cosine s-bends to satisfy the 1.0 dB insertion loss cut-off criterion. This hypothetical gain is represented by the transparent region labelled ‘composite cosine.'

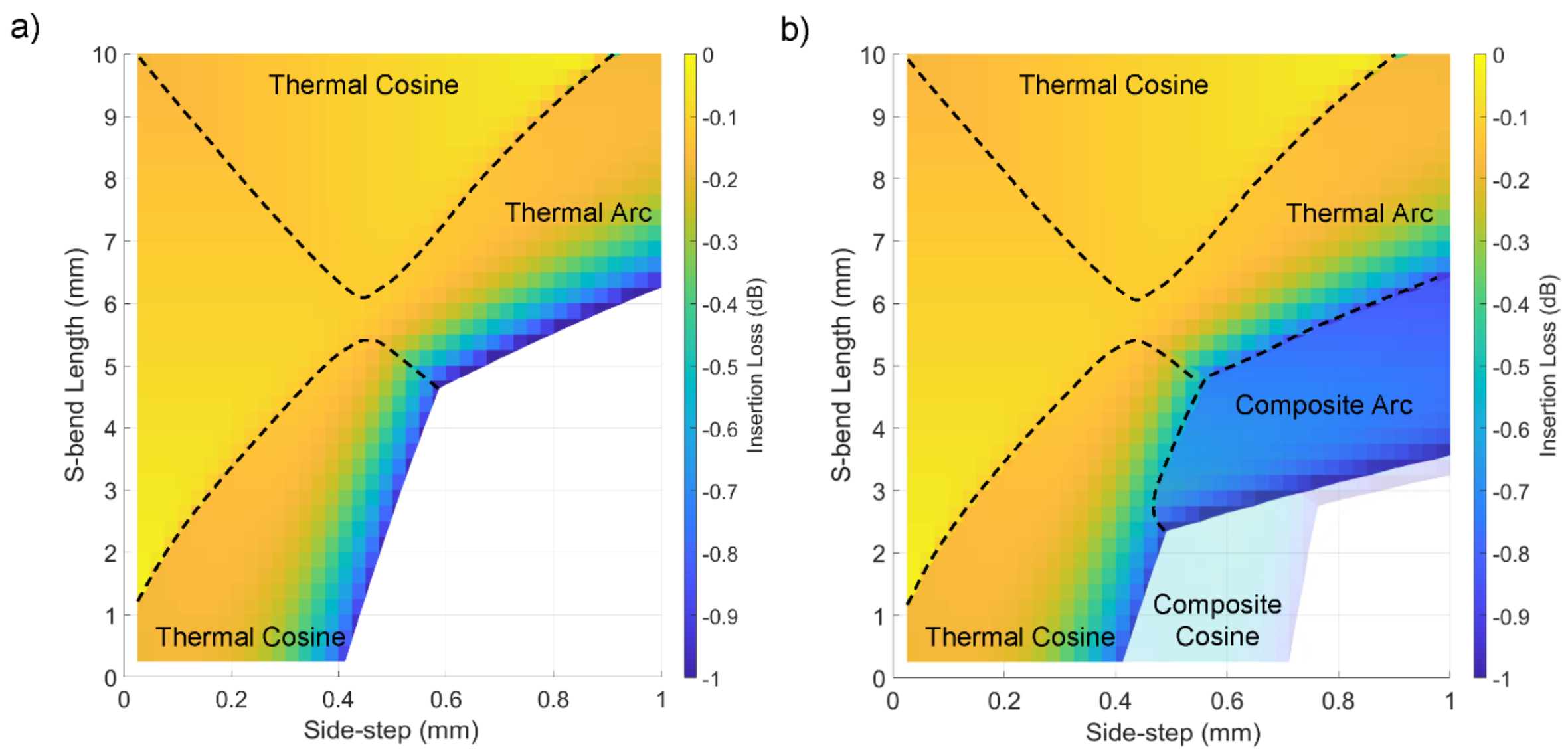

**Figure 4. a) Decision maps illustrating: a) The span of thermal waveguide s-bend parameters for which a 1.0 dB insertion loss criterion is satisfied, and b) the improvement offered by incorporating the composite waveguide technique. The cosine model only applies composite overwriting when the radius of curvature is below 10 mm.**

In summary, we present a novel, composite morphology femtosecond laser written waveguide based on multi-pass inscription featuring both the thermal and athermal modification regimes. The fabrication parameter space was optimised to maximise athermal refractive index contrast, which was found to scale in direct proportion to the pulse energy and linear density (pulses per mm) of the inscription pass, and inverse proportion to the pulse repetition rate. Optimal index contrast was achieved at a repetition rate of 150 kHz, with 200,000 pulses per mm and a pulse energy of 200 nJ. Refractive index change distributions were found to bifurcate; the region of positive index change splits across its central horizontal axis, resulting in two vertically stacked positive lobes with a region of negative index change set between them. This phenomenon occurs above a pulse energy threshold of 110-120 nJ, likely due to the dispersive effects inherent to the focusing of high energy, spectrally broad, short duration pulses. This was addressed by exploring the effects of focusing conditions on the modification morphology, via experimenting with the focusing objective's spherical aberration correction collar. Setting the collar to 0.50 mm, with an inscription depth of 0.25 mm enabled maximal useful peak positive index change, with a magnitude of $1.2 \times 10^{-2}$, measured using QPI at 600 nm. We conclude with demonstration of the composite waveguide and rigorous study of propagation and bend losses for both fabrication regimes. Bend loss analysis provides a very accurate tool for extracting refractive index contrast, allowing for the correction of a 30% discrepancy when measuring at 600 nm versus 1550 nm, using traditional techniques such as RNF. Peak positive index changes of $6.5 \times 10^{-3}$ and $1.12 \times 10^{-2}$ were observed for the thermal and composite waveguide structures, respectively. By combining the thermal and athermal inscription regimes, the 1.0 dB $cm^{-1}$ cut-off for bend radius is reduced from 11 mm to 3.5 mm, a substantive advancement conducive to increased integration density for future PICs such a universal photonic processors and multi-core fibre fan-outs.

## Materials and Methods

### Experimental setup

All waveguides in this work were written in Corning Eagle XG glass, using a Ti:sapphire femtosecond laser (Femtosource XL500, Femtolasers GmbH), which operates at 800 nm and emits 50 fs pulses with a repetition rate of 5.1 MHz. Circularly polarised pulses were focused into the glass at a depth of 250 μm using a 1.4 NA Olympus UPLANSAPO 100× oil immersion microscope objective for thermal modification, and a 0.6 NA Olympus LUCPlanFLN 40× air objective for performing athermal modification. Sample translation through the inscription beam was controlled using an Aerotech A3200 motion controller and 3-axis air bearing linear stages. Additionally, residual stress fields induced by thermal inscription result in a positive halo around the waveguide with a magnitude on the order of $3.0 \times 10^{-3}$. After the first inscription pass, the samples were subjected to a thermal annealing process, as described by *Arriola et al.*[18]. The samples are raised to a temperature of 750 °C, which relaxes the surrounding stress field, removing the halo, and effectively shrinks the waveguide down to the single-mode regime while retaining a high index contrast. Compared to the thermal regime, the athermal regime, by its nature, exhibits reduced thermal stability. The positive index change is the net result of several types of defects in the glass network, the concentrations of which, are differentially modulated by temperature. For example, silicon unpaired bond defects are thermally stable only up to 110 °C. This has been examined in detail by *Fernandez et al.,* who reported less than a 5% reduction in the positive index change for athermal waveguides in fused silica, when annealed at 400 °C (ref. 25).

After annealing, the chip's end-faces are ground back in order to remove waveguide termination artifacts, and then polished to optical surface quality. Thermal waveguides were inscribed with a feedrate of 1500 mm/min, with a pulse energy of 183 nJ, and a pre-annealed width of 30 µm. Waveguides of this size are well characterised, feature low propagation losses of 0.2 dB $cm^{-1}$, good bend loss performance, and circular near-field profiles for low coupling losses to standard telecommunication single-mode fibers on the order of 0.06 dB $cm^{-1}$. Measurement of thermal waveguide refractive index distributions was performed with a Rinck refracted near field (RNF) profilometer[26] (632 nm measurement wavelength); athermal modifications were characterised using a Phasics SID4HR camera (600 nm imaging wavelength), which is capable of quantitative phase imaging (QPI) based on quadriwave lateral shearing interferometry[27] . Mode-field diameters were measured by coupling 1550 nm laser light into the chip using a single-mode optical fibers and imaging the near-field output onto a Xenics Bobcat InGaAs SWIR camera.

**Parameter Study**

Before proceeding with multi-pass fabrication, it was first necessary to optimise the inscription parameters for the secondary athermal pass. The secondary modification is, in essence, a multi-scan waveguide being written inside the core of a thermal waveguide. The athermal modification geometry is sensitive to focusing conditions and strongly confined to the focal volume. Accordingly, the width of the modification is only on the order of 1 μm. Hence, waveguides written in this regime are volumetrically constructed by inscribing multiple closely spaced scanlines in a raster pattern[32]. The parameter space to be explored includes pulse energy (nJ), pulse repetition rate (kHz), pulses per mm, spherical aberration, feedrate (mm/min) and scanline inscription order. Principally, we desired a multiscan waveguide which features high index contrast and symmetric morphology. To this end, the first step of the optimisation was the implementation of the half-scan algorithm which interleaves scan-line inscription order, evenly distributing the localised stress field. This yields a uniform region of index change and also raises the mechanical failure threshold, allowing for higher inscription pulse energies and, thus, higher index contrast modifications[33]. For all experiments the half-scan waveguide width was fixed at 4.8 μm by scan-line placement. This width was selected since it produces a square region of positive index change, and nicely fits inside the core region of a 30 μm (pre-annealed) thermal waveguide. The half-scan modification is comprised of 12 scan-lines, with a minor pitch of 400 nm, and a major pitch of 1200 nm.

**Pulse Repetition Rate and Density**

It was necessary to establish the optimal pulse density for achieving maximum index contrast. Thus, a parameter scan was fabricated at a depth of 250 μm. Pulse energy was fixed at 90 nJ, and the length of the half-scan stub was locked at 12 μm. This pulse energy was selected because it had been observed during QPI analysis that this corresponds to a homogeneous square shaped region of index change. The index contrast was obtained by measuring the optical path length difference across the transverse cross-section and dividing it by the physical length of the stub. The pulse repetition rate was iterated from 150-728 KHz. For each repetition rate, the feedrate was scaled in order to iterate pulse density from 100,000 to 200,000 pulses per mm. The resulting modifications were subjected to QPI, which revealed a positive relationship between refractive index contrast and pulse density, this is shown in Figure 5. a). A slight inverse proportionality was observed between pulse repetition rate and index contrast, allowing for an improvement of approximately 10% at 150 kHz, compared to 728 kHz. For all subsequent experiments, the pulse density was fixed at 200,000 per mm, and the repetition rate was fixed at 150 kHz - the upper and lower bounds, respectively, of the explored parameter space, yielding maximum index contrast. It is possible that contrast could be improved by pushing the envelope further, this is impractical however, as fabrication times become prohibitively long.

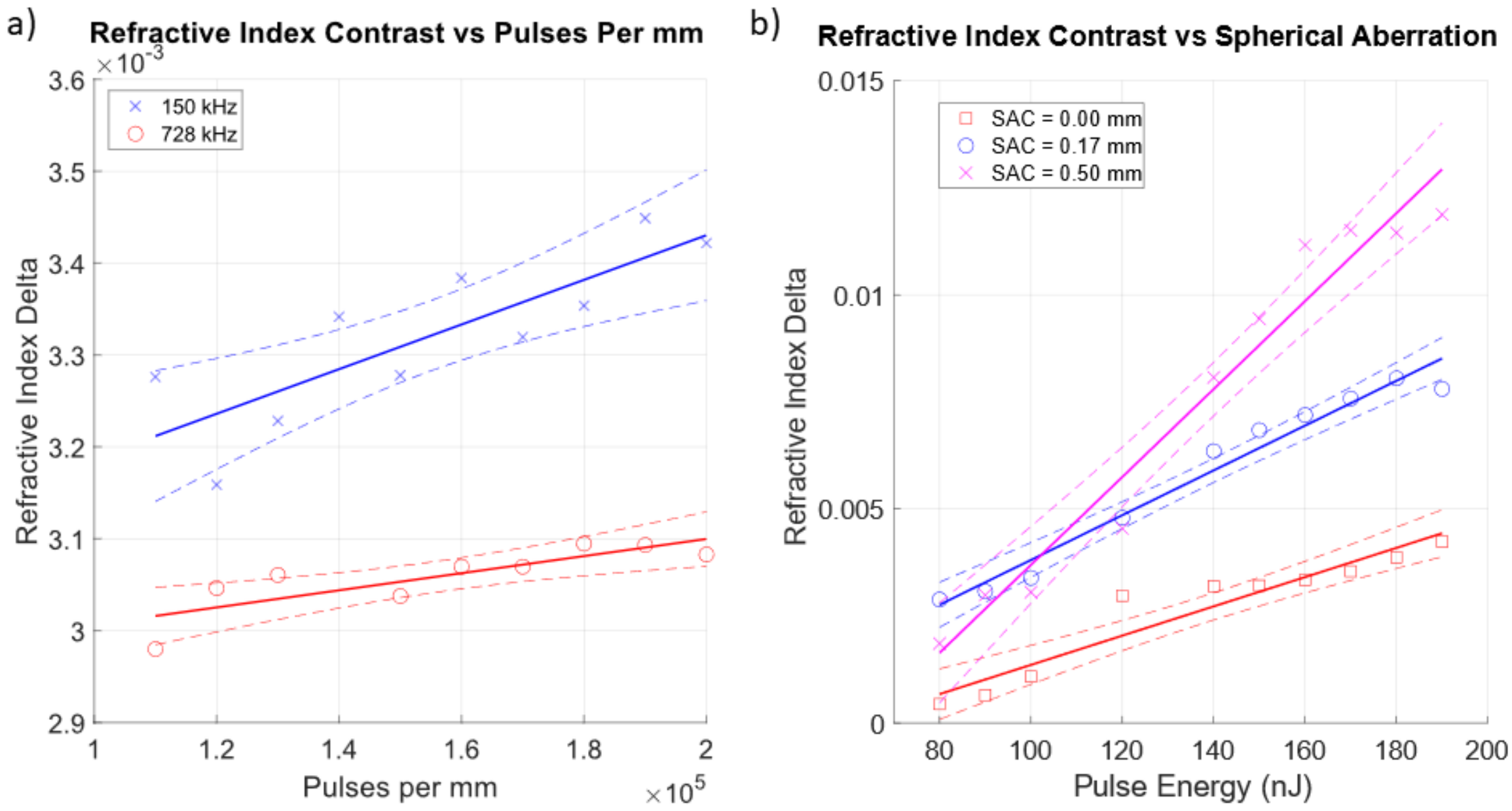

**Figure 5. a) Experimental characterisation of the influence of pulse repetition rate and pulse spatial density on peak positive refractive index change. b) Experimental measurement of peak positive index change as a function of spherical aberration compensation conditions, with a repetition rate of 150 kHz.**

## Spherical Aberration

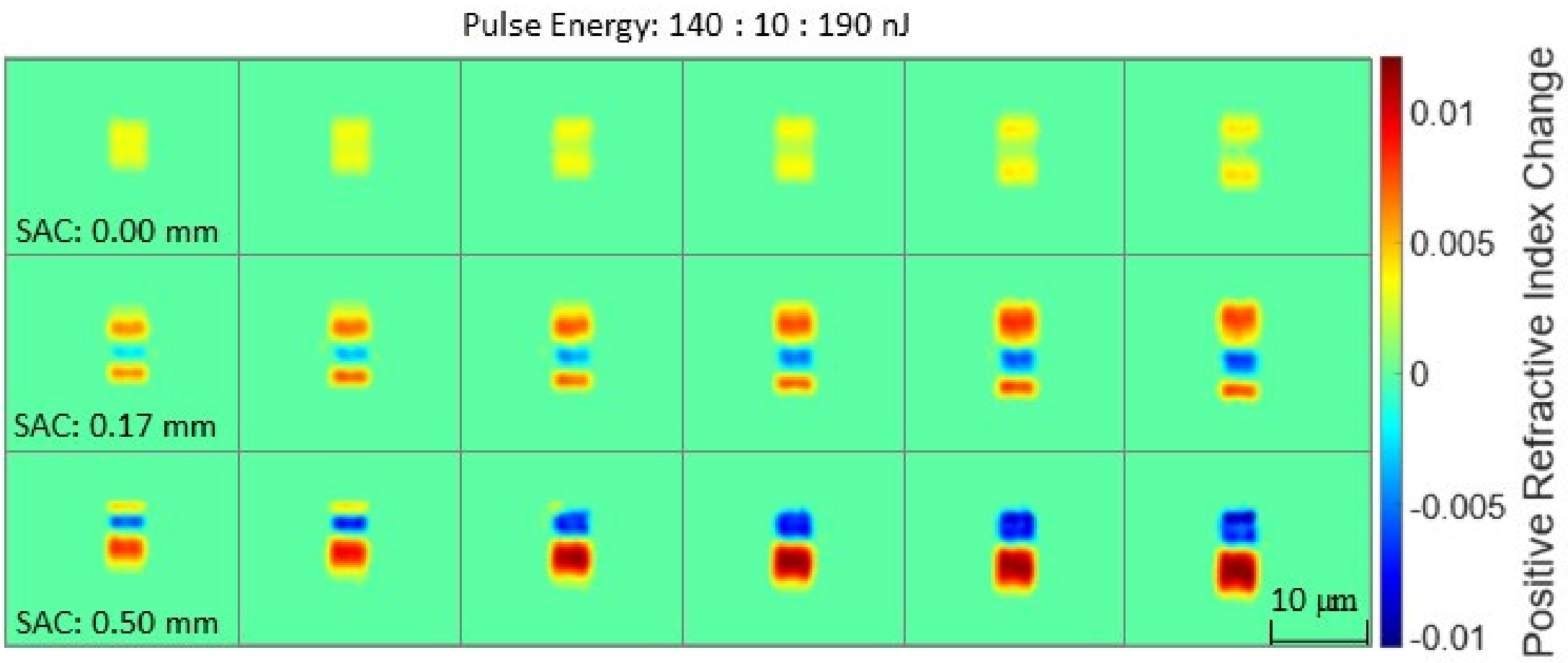

**Figure 6. Collage of quantitative phase imagery of athermal multiscan modifications inscribed at a depth of 250 µm in Eagle XG glass. The evolution of modification morphology as a function of focusing conditions and pulse energy is studied by scanning the power of the incident beam and varying the setting of the focusing objective's spherical aberration collar (SAC).**

It was observed in all cases, that there exists a critical threshold energy at which the region of positive index change bifurcates across its vertical axis, with a negative index change forming between the resulting positive lobes. This energy threshold was utilised as a reference point for future fabrication runs and is hence referred to as the bifurcation energy. It is speculated that the bifurcation phenomenon is a consequence of the short, 50 fs pulse duration, since this equates to a broader spectral bandwidth. As studied by *Sun et al.*[34] this incurs more severe dispersive effects such as temporal broadening of the pulse and wavelength dependent path-length difference which result in non-trivial instantaneous intensity profiles. This bifurcation is problematic since it creates complex guidance conditions subject to multi-mode interference. Various works have investigated the consequences of spherical aberration with regards to both the geometry of the focal volume and the resulting waveguide morphology[35-37], where aberration is controlled by varying the fabrication depth, varying the refractive index of immersion oil, and via the use of spatial light modulators. In this instance, the spherical aberration is varied using the compensation collar built into the focusing objective, which features gradations denoting compensation depths of 0.00 mm, 0.17 mm and 0.50 mm. At a focusing depth of 0.25 mm, when the Spherical Aberration Correction collar (SAC) of the focusing objective is set to 0.00 and 0.17 mm, the bifurcation phenomenon manifests at approximately 110 nJ pulse energy, which effectively caps the maximum useful positive index change at $3.0 \times 10^{-3}$. Figure 5. b) shows the respective refractive index contrast curves with respect to pulse energy, for the different SAC settings. Figure 6. shows a QPI map of the athermal modification morphology with reference to the parameter space being interrogated.

It was discovered that this limitation can be circumvented by setting the SAC to 0.50 mm. Under this regime, the bifurcation phenomenon occurs as normal at 110-120 nJ, however, as the pulse energy continues to increase, the distribution of the index change progressively evolves to yield a single, relatively symmetric region of strong positive index change, located directly below a similarly proportioned region of strong negative index change, as shown in in the bottom row of Figure 6. Consequently, peak positive index change is no longer constrained by the bifurcation limit, allowing positive index changes on the order of $1.12 \times 10^{-2}$, relative to the bulk material, to be readily obtained. The effect of spherical aberration on the relationship between pulse energy and peak positive refractive index change is shown in Figure 5. b). The significance of spherical aberration for boosting the maximum accessible index contrast has also been noted in other recent works[38], and likely was exploited unintentionally during early multiscan demonstrations[39]. It should also be noted that the presence of the strong negative region will serve to compress the mode-field in the vertical axis.

## Alignment

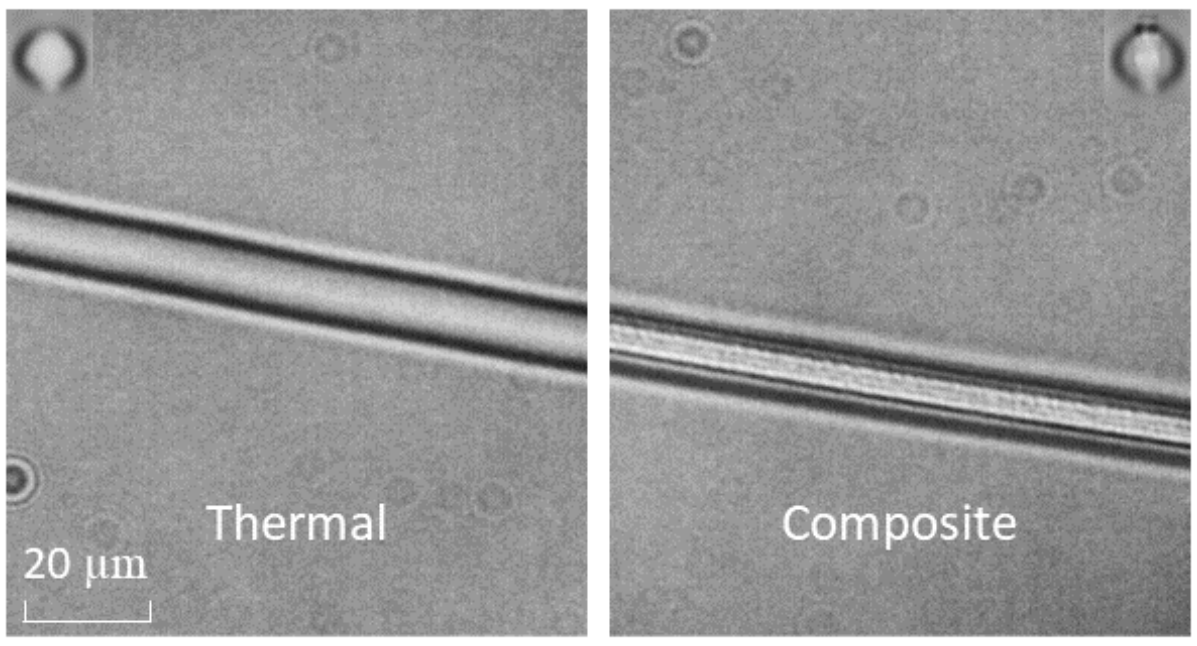

**Figure 7. Bright-field microscope images of a 30 µm pre-annealed width, thermal waveguide, before and after the addition of the athermal inscription pass. Inset images show the corresponding brightfield transverse cross-sections.**

Good alignment of the athermal modification and the thermal waveguide core is critical for minimising bend losses. It was, hence, necessary to develop a reliable recipe for high quality alignment over path lengths up to 8 mm. Alignment tracks, composed of highly visible optical damage, are incorporated into the thermal inscription pass and precisely placed at known positions relative to the thermal waveguides and key coordinates such as the s-bend segment termini. A point-grid height profile is then measured across the surface of the chip, after which, the chip is thermally annealed and polished as described in the methodology section. For the athermal pass, the chip is placed at the same position on the vacuum chuck and the height profile is re-measured to verify that the flatness profile has not significantly deviated. A reference athermal parameter scan is inscribed in close proximity to the chip end-face. The chip is temporarily removed from the fabrication setup for microscope inspection to calibrate pulse energy relative to the bifurcation threshold and fine tune the vertical offset relative to the thermal waveguides. The alignment tracks are then set as the co-ordinate origin and the athermal inscription process begins. Transverse and longitudinal bright-field microscopic inspection of the resulting waveguides was performed to verify quality of the waveguide core overwriting alignment; an example is shown in Figure 7. The alignment recipe proved very successful for a 1 μm of deviation tolerance at any point of the point-grid height profile. Sub-micron lateral alignment was readily attained using the reference markers. The inscription, annealing and post-processing of a photonic chip fabricated in the thermal regime, typically span a period of two days. The addition of the multi-pass step adds an additional half-day of processing time.

**Acknowledgements**

This work is funded by the Australian Research Council Discovery Program under grant FT200100590. The work was performed in part at the OptoFab node of the Australian National Fabrication Facility (ANFF) utilizing Commonwealth as well as NSW state government funding.

**Author Contributions**

S.G. supervised and resourced the project. A.J.R.A, T.F. and S.G. conceived the experiments. A.J.R.A carried out design, fabrication and execution of the experiments. All authors participated in the analysis of data and contributed to the writing of manuscript.

**Conflict of interest**

The authors declare no competing interests.

**Supplementary information**

Supplementary materials are available on request.